\documentstyle[12pt,epsfig]{article}
\begin{document}
\noindent
\begin{center}
{\Large {\bf Inflation and Cosmological Constant  }}\\
\vspace{2cm}
 ${\bf Yousef~Bisabr}$\footnote{e-mail:~y-bisabr@sru.ac.ir.}\\
\vspace{.5cm} {\small{Department of Physics, Shahid Rajaee Teacher
Training University,
Lavizan, Tehran 16788, Iran}}\\
\end{center}
\begin{abstract}
In construction of an inflationary model, one usually assumes that the matter sector of the gravitational action is minimally coupled to the background. It means that the matter (inflaton) part of the action is coupled with the same metric of the gravitational part. We elaborate on this  assumption and investigate some of the consequences. We assume that the gravitational and the inflaton sectors belong to different units (or conformal frames).
We show that this coupling can convert a single-field inflationary model into a two-field one with a mixed kinetic term. The energy-momentum tensor of the inflaton is therefore non-conserved due to the interaction with the conformal factor. This allows an energy exchange between the two fields and provides us with a mechanism for reduction of a large effective cosmological constant during inflation.
 \end{abstract}

~~~~~~~~~~~~~~~~~~~~~~~~~~~~~~~~~~~~~~~~~~~~~~~~~~~~~~~~~~~~~~~~~~~~~~~~~~~~~~~~~~~~~~~~~~~~~~~~~~~~~
\section{Introduction}
Standard cosmology (the Big Bang cosmological model) is the most successful model in describing the observed Universe. It explains the redshifts observed in spectral lines of all galaxies, the origin of Cosmic Microwave Background (CMB) Radiation, abundances of light elements and formation of large structures. Despite these successes, there are unresolved problems such as flatness \cite{flat} and horizon \cite{hor} problems. The former concerns with the fact that the amount of matter in the Universe is just sufficient to halt its expansion. The latter, on the other hand, arises due to absence of a mechanism that leads to the same initial conditions in causally disconnected regions of space. Both these problems are related to fine tuning of initial conditions at early times.\\
Another problem of the standard cosmology is the cosmological constant (CC) problem \cite{wein}. It is generally believed that CC represents a gravitational contribution of vacuum fluctuations to Einstein equations in General Relativity \cite{bis1}.
According to particle physics estimations, the energy density associated to vacuum is $\rho_{\Lambda}^{th}\equiv \frac{\Lambda_{th}}{8\pi G}\sim(10^{18}Gev)^4\sim 10^{110}erg/cm^3$ \cite{th}. This conflicts with recent observations which suggest that $\rho_{\Lambda}^{obs}\equiv \frac{\Lambda_{obs}}{8\pi G}\sim(10^{-12}Gev)^4\sim 10^{-10}erg/cm^3$  \cite{obs} \cite{ccc}. The CC problem is the sharp contradiction between
theory and observations which is an incredible 120 orders of magnitude. \\
One can find an explanation for the flatness and horizon problems if one assumes that the Universe has been passed
through an accelerated expansion phase at early times.
There are various types of models which generate such an inflationary phase. In a simple single-field model, one uses a scalar field, the so-called inflaton, described by a Lagrangian density such as
\begin{equation}
L(g_{\mu\nu},
\phi)=\frac{1}{2}g^{\mu\nu}\nabla_{\mu}\phi\nabla_{\nu}\phi+V(\phi)
\label{0-2}\end{equation} The function $V(\phi)$ is the
corresponding potential which cotrols the dynamics of $\phi$. This Lagrangian density together with the
Einstein-Hilbert term give the total inflation action\footnote{We
work in the unit system in which $\hbar=c=1$.}
\begin{equation}
S=\frac{1}{2\kappa^2} \int d^{4}x \sqrt{-g} R -\int d^{4}x \sqrt{-g}
L(g_{\mu\nu}, \phi)\label{0-1}\end{equation}
 where $\kappa^2=8\pi M_p^{-2}$ with $M_p$ being the Planck mass. For a spatially flat
Friedmann-Robertson-Walker metric
\begin{equation}
ds^2=-dt^2+a^2(t)(dx^2+dy^2+dz^2)
\label{frw}\end{equation}
the time evolution of the scale factor $a(t)$ and the inflaton $\phi(t)$ are governed by
\begin{equation}
3H^2=\kappa^2(\frac{1}{2}\dot{\phi}^2+V(\phi))
\label{0-3}\end{equation}
\begin{equation}
\ddot{\phi}+3H\dot{\phi}=-V'(\phi) \label{0-4}\end{equation} where
$H=\frac{\dot{a}}{a}$ and the overdot and the prime indicate differentiation
with respect to $t$ and $\phi$, respectively.  In order to have an inflationary phase, the potential energy in the right hand side of (\ref{0-3}) should dominate over the kinetic energy. This is possible provided that the potential is flat enough so that the scalar field
slowly rolls down the potential $V(\phi)$. In this slow-roll
approximation
$\frac{1}{2}\dot{\phi}^2<<V(\phi)$ or $\phi\approx const$ and then (\ref{0-3}) and (\ref{0-4}) become
 \begin{equation}
3H^2\approx\kappa^2 V(\phi)
\label{0-33}\end{equation}
\begin{equation}
3H\dot{\phi}\approx -V'(\phi) \label{0-44}\end{equation}
leading to an exponential expansion (a de sitter solution) for the scale
factor. If we define slow-roll parameters
\begin{equation}
\epsilon_{V}(\phi)=\frac{M_p^2}{2}(\frac{V'(\phi)}{V(\phi)})^2
\label{0-5}\end{equation}
\begin{equation}
\eta_{V}(\phi)=M_p^2(\frac{V''(\phi)}{V(\phi)})
\label{0-6}\end{equation}
where the first measures the slope of the potential and the second the curvature,
then necessary conditions for the slow-roll approximation to hold are $\epsilon_V<<1$ and $|\eta_V|<<1$.\\
Two important aspects concerning the above procedure should be emphasized at this point. Firstly, it
suffers from a severe fine-tuning problem, namely the shape of the potential $V (\phi)$ should be quite flat in order to have enough inflation. In order to remove this deficiency, chaotic inflation was introduced based
on evolution of different possible distributions of a scalar field
without making ad hoc assumptions on the shape of the potential
function \cite{lin1}. Secondly, even though inflationary scheme is based on a large CC (vacuum energy density of the inflaton), it provides no explanation for the CC problem. Here we intend to modify the inflation action (\ref{0-1}) based on simple physical principles in order to alleviate the CC problem during inflation.

~~~~~~~~~~~~~~~~~~~~~~~~~~~~~~~~~~~~~~~~~~~~~~~~~~~~~~~~~~~~~~~~~~~~~~~~~~~~~~~~~~~~~~~~~~~~~~~~~~~~~~~~~~~~~~~~~~~~~~~~~~~~~~
\section{A model of inflation}
We consider two units systems $U_G$ and $U_A$ which are defined in terms of fundamental constants usually used in gravitational theories and atomic (quantum) physics. Thus they may be called the gravitational and the atomic units systems. A spacetime interval $dS$ which is measured in these units are related by $dS_{G}=\beta~ dS_A$ which $\beta$ is the conversion factor. In every day life (or in the human time scale) $\beta$ is taken to be a constant which means that the two units $U_G$ and $U_A$ are constant multiple of each other. However, there are arguments against the constancy of $\beta$ in a time scale comparable to the age of the Universe \cite{bis1}. Following these arguments, we investigate the consequences of the following assumption
\begin{equation}
dS_G=\beta(t)~ dS_A
\label{1-88}\end{equation}
where the scale function $\beta(t)$ is now a function of the epoch. In this viewpoint, a units transformation should be regarded as a dynamical process rather than a naive multiplication by a
constant factor. If the metrics corresponding to $dS_G$ and $dS_A$ are $g_{\mu\nu}$ and $\bar{g}_{\mu\nu}$, respectively,
\begin{equation}
dS_G^2=g_{\mu\nu} dx^{\mu}dx^{\nu}
\end{equation}
\begin{equation}
dS_A^2=\bar{g}_{\mu\nu} dx^{\mu}dx^{\nu}
\end{equation}
 then (\ref{1-88}) will be equivalent to a conformal transformation
\begin{equation}
\bar{g}_{\mu\nu}=\beta^{-2}(t)g_{\mu\nu}
\end{equation}
We now apply this viewpoint to the simple inflationary model (\ref{0-1}). We would like to consider a generalization of gravitational coupling of inflaton field beyond coupling to the background geometry. Denoting the background metric and inflaton scalar field by $(g_{\mu\nu}, \phi$), we will consider the case that the whole inflaton sector of an inflationary model to be described by the conformally re-scaled fields ($\bar{g}_{\mu\nu}, \bar{\phi}$) where $\bar{\phi}=\beta\phi$. In this case, the model of inflation (\ref{0-1}) becomes
\begin{equation}
S=\frac{1}{2\kappa^2} \int d^{4}x \sqrt{-g} R -\int d^{4}x
\sqrt{-\bar{g}} L(\bar{g}_{\mu\nu},\bar{\phi})
\label{1-4}\end{equation}
where
\begin{equation}
L(\bar{g}_{\mu\nu},
\bar{\phi})=\frac{1}{2}\bar{g}^{\mu\nu}\nabla_{\mu}\bar{\phi}\nabla_{\nu}\bar{\phi}+V(\bar{\phi})
\label{1-1}\end{equation}
Physical interpretation of this coupling of the inflaton field relies on the interpretation of conformal transformations as local units transformations, namely that transformations with a spacetime-dependent conversion factors \cite{bek2}. In this sense, one can say that the two sectors of (\ref{1-4}) are given in two different unit systems.\\
We take $\beta=e^{-\lambda\sigma}$ where $\sigma$ is a dimensionless smooth function and $\lambda$ as a constant parameter. Then we write
 the action (\ref{1-4}) in terms of the background variables ($g_{\mu\nu}, \phi$)
\begin{equation}
S=\frac{1}{2} \int d^{4}x \sqrt{-g}~ \{\frac{1}{\kappa^2}R-g^{\mu\nu}
\nabla_{\mu} \phi \nabla_{\nu}\phi+2\lambda\phi g^{\mu\nu} \nabla_{\mu} \phi
\nabla_{\nu}\sigma-\lambda^2\phi^2 g^{\mu\nu} \nabla_{\mu}
\sigma \nabla_{\nu}\sigma-2V(e^{\sigma}\phi)e^{4\lambda\sigma}\}
\label{1-6}\end{equation}
In this representation, the inflationary model appears as a two-field model with a kinetic mixing term (the third term in the above action) \cite{mix}.
The field equations resulting from (\ref{1-6}) are
\begin{equation}
G_{\mu\nu}=\kappa^2(T_{\mu\nu}+\tau_{\mu\nu})
\label{ein}\end{equation}
\begin{equation}
\Box\phi-\lambda\phi\Box\sigma-\lambda^2\phi\nabla_{\gamma}\sigma\nabla^{\gamma}\sigma-U_{\phi}=0
\label{pp}\end{equation}
\begin{equation}
\lambda^2\phi^2\Box\sigma-\lambda\phi\Box\phi-\lambda\nabla_{\gamma}\phi\nabla^{\gamma}\phi
+2\lambda^2\phi\nabla_{\gamma}\phi\nabla^{\gamma}\sigma-U_{\sigma}=0
\label{s}\end{equation}
where $U(\phi, \sigma)\equiv V(\phi, \sigma)e^{-4\sigma}$, $U_{\phi}\equiv\frac{\partial U}{\partial\phi}$, $U_{\sigma}\equiv\frac{\partial U}{\partial\sigma}$ and
\begin{equation}
T_{\mu\nu}=(\nabla_{\mu}\phi\nabla_{\nu}\phi-\frac{1}{2}g_{\mu\nu}\nabla_{\gamma}\phi\nabla^{\gamma}\phi)
-Ug_{\mu\nu}
\end{equation}
\begin{equation}
\tau_{\mu\nu}=\phi^2(\nabla_{\mu}\sigma\nabla_{\nu}\sigma-\frac{1}{2}g_{\mu\nu}\nabla_{\gamma}\sigma\nabla^{\gamma}\sigma)
+2\phi(\nabla_{\mu}\phi\nabla_{\nu}\sigma-\frac{1}{2}g_{\mu\nu}\nabla_{\gamma}\phi\nabla^{\gamma}\sigma)
\label{ta}\end{equation}
 We note that when $\sigma$ takes a constant configuration, $\tau_{\mu\nu}=0$ and the equations (\ref{ein}) reduce to the usual Einstein equations in General Relativity.

~~~~~~~~~~~~~~~~~~~~~~~~~~~~~~~~~~~~~~~~~~~~~~~~~~~~~~~~~~~~~~~~~~~~~~~~~~~~~~~~~~~~~~~~~~~~~~~~~~~~~~~~~~~~~~~~~~~~~~~~~~
\section{Cosmological setting}
For interpreting the above set of equations in a cosmological setting, we write them for the metric (\ref{frw}) and obtain
\begin{equation}
3H^2=\kappa^2(\frac{1}{2}\dot{\phi}^2+\frac{1}{2}\lambda^2\phi^2\dot{\sigma}^2-\lambda\phi\dot{\phi}\dot{\sigma}+U)
\label{1-12}\end{equation}
\begin{equation}
\dot{H}+H^2=-\kappa^2(\frac{1}{3}\dot{\phi}^2+\frac{1}{3}\lambda^2\phi^2\dot{\sigma}^2-\frac{2}{3}\lambda\phi\dot{\phi}\dot{\sigma}
-\frac{1}{3}U)
\label{1-13}\end{equation}
\begin{equation}
\ddot{\phi}-\lambda\phi\ddot{\sigma}+3H(\dot{\phi}-\lambda\phi\dot{\sigma})-\lambda^2\phi\dot{\sigma}^2+U_{\phi}=0
\label{1-14}\end{equation}
\begin{equation}
\lambda\phi(\lambda\phi\ddot{\sigma}-\ddot{\phi})+3\lambda H\phi(\lambda\phi\dot{\sigma}-\dot{\phi})+\lambda\dot{\phi}(2\phi\dot{\sigma}-\dot{\phi})+U_{\sigma}=0
\label{1-15}\end{equation}
In a two-field inflationary model, we need to apply the slow-roll scheme to the both scalar fields. To do this, we neglect the terms containing second time derivatives of the fields in the above equations and obtain
\begin{equation}
3H^2\simeq\kappa^2 U
\label{1-12a}\end{equation}
\begin{equation}
\dot{H}+H^2\simeq \frac{1}{3}\kappa^2 U
\label{1-13a}\end{equation}
\begin{equation}
\dot{\phi}\simeq \frac{-U_{\phi}}{3H}+\lambda\phi\dot{\sigma}
\label{1-14a}\end{equation}
\begin{equation}
3\lambda H\phi(\lambda\phi\dot{\sigma}-\dot{\phi})\simeq -U_{\sigma}
\end{equation}
It is interesting to compare these slow-roll equations with those of single-field models.
The first two equations indicate existence of a de sitter phase ($H\simeq const$).
One important feature is related to (\ref{1-14a}). This equation reduces to the usual relationship between $\dot{\phi}$ and $U_{\phi}$ in the slow-roll approximation only for $\sigma=const$. For $\dot{\sigma}\neq 0$, the rolling-velocity $\dot{\phi}$ depends not only on the shape of the potential but on the dynamics of $\sigma$. In this case, $\dot{\phi}$ is smaller ($\lambda > 0$) or larger ($\lambda < 0$) than
the rolling-velocity of $\phi$ in the case that $\dot{\sigma}=0$.
Moreover, due to the interaction of $\phi$ and $\sigma$ there is an energy transfer between these energy components. Actually, (\ref{ein}) indicates that $\nabla^{\mu}T_{\mu\nu}\neq 0$
 which means that $T_{\mu\nu}$ is not conserved because of the interaction between $\phi$ and $\sigma$. Due to this interaction there is an energy transfer between these energy components. To show this more explicitly, let us consider the case that $\lambda\ll 1$\footnote{Magnitude of such a coupling parameter is severely constrained by equivalence principle experiments and late-time cosmological observations \cite{late1} \cite{late2}. }. In this case, one may neglect the terms of orders greater than $\lambda^2$ and write
$$
-\nabla^{\mu}T_{\mu\nu}=\nabla^{\mu}\tau_{\mu\nu}\approx -\lambda\Big\{2\nabla^{\gamma}\phi\nabla_{\gamma}\phi\nabla_{\nu}\sigma-
\nabla_{\nu}\phi\nabla_{\gamma}\phi\nabla^{\gamma}\sigma+2\phi\Box\phi\nabla_{\nu}\sigma+\phi\nabla_{\gamma}\phi\nabla^{\gamma}\phi\nabla_{\nu}\sigma
$$
\begin{equation}
-\phi\nabla_{\nu}\nabla_{\gamma}\phi\nabla^{\gamma}\sigma\Big\}~~~~~~~~~~~~~~~~~~~~~~~
\end{equation}
In a homogeneous and isotropic Universe, this corresponds to
\begin{equation}
\dot{\rho}_{\phi}+3H\rho_{\phi}\approx \lambda\phi H \dot{\phi}\dot{\sigma}+ \emph{terms containing third time derivatives}
\label{bal}\end{equation}
where $\rho_{\phi}$ is energy density of the field $\phi$. It is now evident that when $\lambda>0$ ($\lambda<0$), the field $\phi$ gains (loses) energy. The energy loss corresponding to $\lambda<0$ can be taken as a basis for vacuum decay during inflation.\\
As pointed out earlier, the CC problem is related to an extreme confliction between the values of $\Lambda_{th}$ and $\Lambda_{obs}$. One explanation for such a confliction is that a reduction mechanism has been worked during expansion of the Universe by which the large cosmological term is reduced to the present value \cite{bra1}. In construction of such a reduction mechanism one should accept that $\Lambda_{th}$ and $\Lambda_{obs}$ belong to two different epochs of
evolution of the Universe \cite{bis1} \cite{biss}. There are at least two reasons why $\Lambda_{th}$ should be related to the
early times. The first reason concerns with the fact that all theoretical predictions of $\Lambda$ are
based on
fluctuations of quantum fields which are important only at early times where energy
of the fields are so high that classical descriptions are no longer valid. Furthermore, standard
model of particle physics implies that the universe has undergone a series of phase transitions
at early epoch of its evolution leading to injection of a huge energy density into vacuum.\\
The
second reason is related to the inflationary Universe scenario. Inflation
postulates that there has been a phase of exponential expansion at early Universe driven by
a large cosmological constant. Thus a large cosmological constant is very important, at least for early versions of inflationary models, and
whatever physical process that has led to $\Lambda\approx 0$ today must also allow it to take a large value
in the past.\\
The generalization of coupling of the inflaton field proposed here provides a reduction mechanism working at the inflation period. This is evident by noting the exponential pre-factor in $U(\phi,\sigma)=V(\bar{\phi})e^{4\lambda\sigma}$ which appears in (\ref{1-6}) and the field equations (\ref{1-12})-(\ref{1-15}). When $\lambda<0$, $V(\bar{\phi})$ (which may receives contributions from various mass scales
introduced by standard model fields) decays in the course of expansion of the Universe. This interpretation is supported by the balance equation (\ref{bal}) which indicates an energy flow away from $\phi$ field for $\lambda<0$.

~~~~~~~~~~~~~~~~~~~~~~~~~~~~~~~~~~~~~~~~~~~~~~~~~~~~~~~~~~~~~~~~~~~~~~~~~~~~~~~~~~~~~~~~~~
\section{Conclusions}
We have investigated generalization of gravitational coupling of the inflaton in a single-field inflationary model. In this generalization, the gravitational and the inflaton sectors of the total action are given in two different units systems\footnote{The idea that the gravitational and the atomic units may be related by an epoch-dependent conversion factor was originally suggested by Dirac \cite{dir}. He also used two different metrics in Einstein equations to bring them into accord with possible variations of $G$ \cite{bis1}.}.
This
converts the single-field inflationary model into a two-field one (the original inflaton field $\phi$ and the conformal factor $\sigma$)\footnote{It is possible to relate such a two-field model to cosmological parameters in a perturbative level \cite{wands}.} with a mixed kinetic term. The main result of this analysis is the following:\\
Due to the interaction between the inflaton $\phi$ and the field $\sigma$, there is an energy exchange between the two fields. When $\lambda<0$, the energy transfer is out of $\phi$. In this case, the interaction (which is a result of the non-minimal coupling of the inflaton field) provides a mechanism for reducing a large effective cosmological constant. This energy evacuation is also apparent from the exponential damping pre-factor in front of the mass of the inflaton $V e^{-4\sigma}$. It is therefore possible to generalize gravitational coupling of the inflaton in the inflationary scheme so that inflation not only addresses the flatness and the horizon problems but it also alleviates the CC problem. \\\\

{\bf Acknowledgment}\\\\
This work is supported by the Iran National Science Foundation (INSF) with Grant No.6074.

~~~~~~~~~~~~~~~~~~~~~~~~~~~~~~~~~~~~~~~~~~~~~~~~~~~~~~~~~~~~~~~~~~~~~~~~~~~~~~~~~~~~~~~~~~~~~~~

\end{document}